\documentclass[runningheads]{llncs}
\usepackage[T1]{fontenc}
\usepackage{graphicx}
\usepackage{cite}
\usepackage{amsmath,amssymb,amsfonts}
\usepackage{algorithm}
\usepackage{algpseudocode}
\usepackage{graphicx}
\usepackage{textcomp}
\usepackage{xcolor}
\usepackage{booktabs}
\usepackage{microtype}
\usepackage[T1]{fontenc}
\usepackage{lmodern}
\begin{document}
\title{A DQN-based model for intelligent network selection in heterogeneous wireless systems}
%
%
\author{Fayssal Bendaoud\inst{1} \and
Asma Amraoui\inst{2} \and
Karim Sehimi\inst{1}}
\authorrunning{F. Bendaoud et al.}
%
\institute{LabRI-SBA., Ecole Superieure en informatique, Sidi Bel Abbes, Algeria 
\email{f.bendaoud@esi-sba.dz}\\
\and
Laboratory of Telecommunication Tlemcen LTT, University of Tlemcen, Ageria\\
}
\maketitle              
\begin{abstract}
Wireless communications have been at the center of the revolution in technology for the last few years.  The 5G communication system is the pinnacle of these technologies; however 4G LTE, WiFi, and even satellite technologies are still employed worldwide.  So, the aim of the next generation network is to take advantage of these technologies for the better of the end users.  Our research analyzes this subject and reveals a new and intelligent method that allows users to select the suitable RAT at each time and, therefore, to switch to another RAT if necessary.  The Deep Q-Network (DQN) algorithm was utilized, which is a reinforcement learning algorithm that determines judgments based on antecedent actions (rewards and punishments).  The approach exhibits a high accuracy, reaching 93\%, especially after a given number of epochs (the exploration phase), compared to typical MADM methods where the accuracy doesn't exceed. 75\%

\keywords{DQN \and MADM \and network selection \and RAT \and 5G \and 4G \and LTE \and WiFi.}
\end{abstract}
\section{Introduction}
\begin{figure*}[ht]
\centerline{\includegraphics[width=12cm, height=6cm]{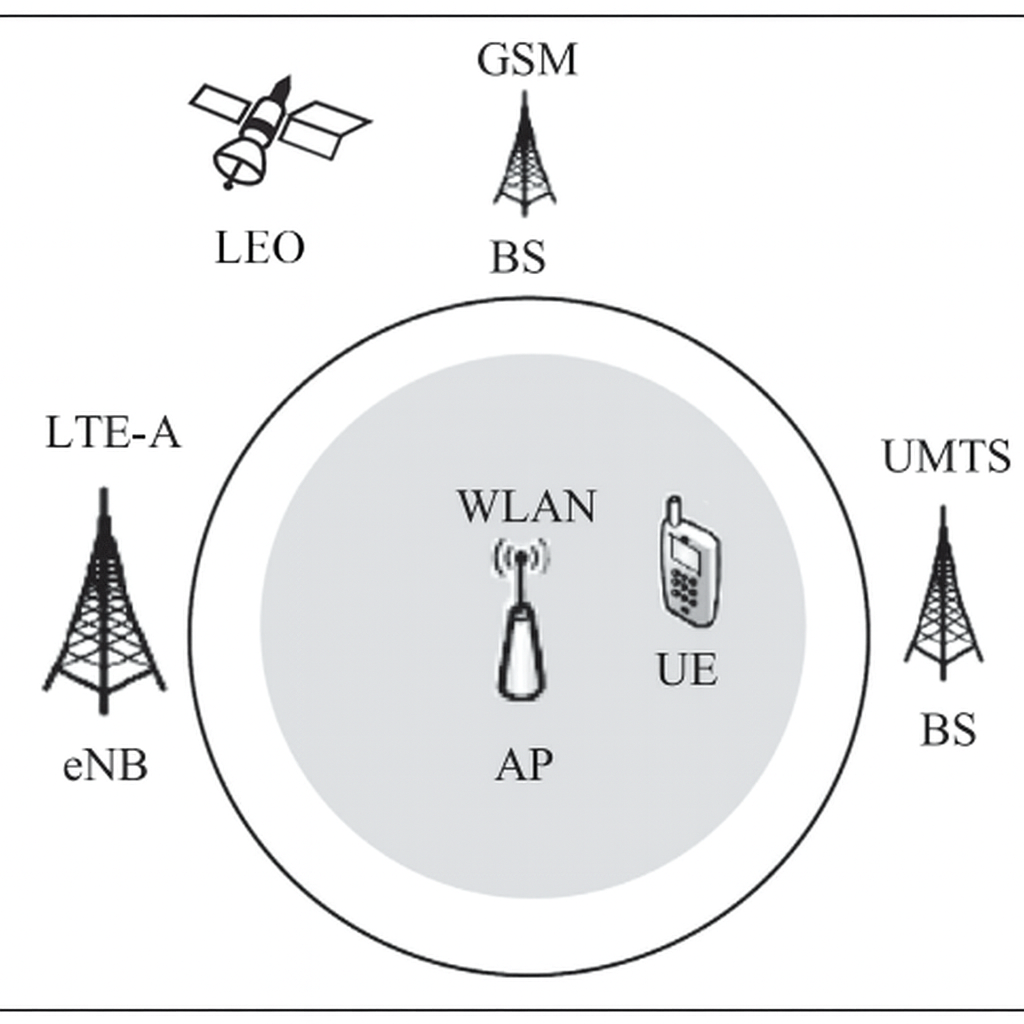}}
\caption{Access network selection in a heterogeneous wireless environment. \label{fig11}\cite{}}
\end{figure*}

The previous decade was characterised by the swift and significant advancement of wireless communication, enabling RATs to provide very high data rates, low latency, little jitter, and diminished packet loss.  The various RATs, each providing distinct QoS (Quality of Service), established what is known as a heterogeneous wireless environment.  In a heterogeneous wireless environment, various Radio Access Technologies (RATs) such as Wi-Fi, 4G LTE, and 5G (specifically eMMB for general users, excluding URLLC and mMTC which cater to autonomous vehicles and IoT systems, respectively), coexist; selecting the optimal network is essential for a seamless user experience. Access network selection, \cite{bendaoud2018management} entails a sophisticated process whereby mobile devices identify ideal networks based on signal strength, bandwidth, and particularly Quality of Service (QoS).  This selection gets exceedingly complicated in heterogeneous situations with dramatically variable network properties that fluctuate wildly under various propagation conditions and RAT loads.  In this context, modern consumers, equipped with multimodal terminals capable of connecting to different networks simultaneously, see Figure \ref{fig11}, want service continuity that translates into a seamless transition between networks. \\
Innovative approaches based on predictive models have been developed to increase the accuracy and efficiency of the network selection process.  For example, the use of Markov models \cite{bendaoud2020mobility} or machine learning algorithms such as KNN (K-nearest neighbours) \cite{bendaoud2021adaptive} allows predicting user movements and dynamically adjusting the network choice according to the anticipated conditions.  Fuzzy logic combined with game theory in decision-making processes provides more effective management of uncertainties and conflicting objectives that unexpectedly appear when integrating the quality of service with energy consumption or usage costs\cite{evangeline2022two}, \cite{sun2022game}.  Ideally, picking access networks in diverse wireless settings \cite{bendaoud2014network} considerably improves the user experience and encourages a more efficient utilisation of resources. \\
 In this research, we are proposing a novel and innovative strategy for this purpose, which consists of employing the reinforcement learning technique, where the idea behind it is that the agent learns from its environment. The agent conducts actions at each time and receives rewards in return; rewards are positive or negative, dependent on its choices, and with more epochs, it will be efficient, and it learns how to select excellent actions that yield positive rewards.  Specifically, we are using the Deep Q Network DQN algorithm in this paper, which is an algorithm combining the Q learning algorithm with a deep neural network.\\
Modern communication systems management relies largely on this process at its heart, playing a crucial part in moving towards next-generation networks, where reacting swiftly under often changing circumstances becomes vital for wireless services continuity. \\
So, we provide a new DQN-based model with dynamic incentive design and adaptability to multiple RATs (WiFi, 4G, 5G, and LEO). Our work includes a thorough comparison of MADM techniques and an evaluation of learning behaviour in terms of exploration-exploitation trade-offs. \\
This paper is organised as follows:  in part 2, we give a review of the literature on models for the selection of the access network in a heterogeneous wireless environment; in part 3, the formulation of the problem, the presentation of the different algorithms utilised in this study, and finally, the presentation of our approach.  Part 4 is dedicated to the results and their interpretation, and we conclude the work with a viewpoint conclusion.
\section{Related work}
The selection of access networks in heterogeneous wireless networks (HWN) has been extensively addressed utilising numerous methodologies and approaches, including Multiple Attribute Decision Making (MADM), objective functions, evolutionary algorithms, fuzzy game theory, and machine learning. The objective was always to maximise the customer experience and service quality. These techniques examine several aspects, including user preferences, network circumstances (load and congestion), and service characteristics \cite{bendaoud2018management}, \cite{wang2012mathematical}. In the following, we will present a comprehensive review of these numerous strategies employed in the situation of access network selection.\\
The paper \cite{bendaoud2017modified} introduces a modified-SAW technique; the authors conducted comparisons using several Multi-Attribute Decision-Making (MADM) methodologies. The results indicate that the modified SAW outperforms the alternatives, even in the situation of rank reversal, which adversely affects all MADM approaches, but their solution circumvents this issue. \\ 
Another study \cite{bendaoud2020mobility} by the same authors employs an objective function aimed at minimising the frequency of handoffs for mobile users by using a Markov process for the path prediction part. They claim that the study offers outcomes even better than MADM approaches. \\
Other techniques, such as game theory, were used, as referenced in \cite{bendaoud2015network} and \cite{salih2015novel}. Objective functions were used, as shown in \cite{nguyen2008utility} and \cite{yu2019hybrid}; moreover, fuzzy logic was utilised for this purpose, as indicated in \cite{evangeline2022two}.  \\
Machine learning algorithms were utilised to cope with the access network selection challenge. The article \cite{bendaoud2024machine} provides a technique that outperforms standard multiattribute decision-making approaches by minimising latency and packet loss, guaranteeing that users are "always best connected". \\ 
 In addition, the work \cite{bendaoud2021adaptive} addressed the same issue, and the author suggested a modified version of the classic KNN algorithm that permits the RAT ranking.  The author believes that it delivers excellent results compared to standard procedures. \\ 
 The authors in \cite{wang2019intelligent} proposed a mix between game theory and machine learning to achieve fast and optimal network selection; the implementation of their intelligent algorithm was at the user side; they claim that the simulations confirmed the effectiveness of the proposed algorithm in reducing frequent switching, reducing average delay, enhancing user experience, and increasing resource utilisation. \\ In \cite{de2022satellite}, the authors describe a DQN (Deep Q-Network) algorithm.  They confirm that they are employing a realistic scenario that involves route loss models and intra-cell interference.  They claim that with this technique, they increase load balancing, the utilisation of resources, and link approval. \\ 
 A deep reinforcement learning framework was presented to optimise the selection across heterogeneous health systems in \cite{chkirbene2021deep}.  They claim that their solution provides for the decrease of the transmitted energy, the monetary cost, and distortion while preserving the overall QoS.  They further state that, compared to greedy methods, their system is considerably superior and can adapt to dynamic changes. 
 \section{Access network selection problem}
\begin{figure*}[ht]
\centerline{\includegraphics[width=12cm, height=6cm]{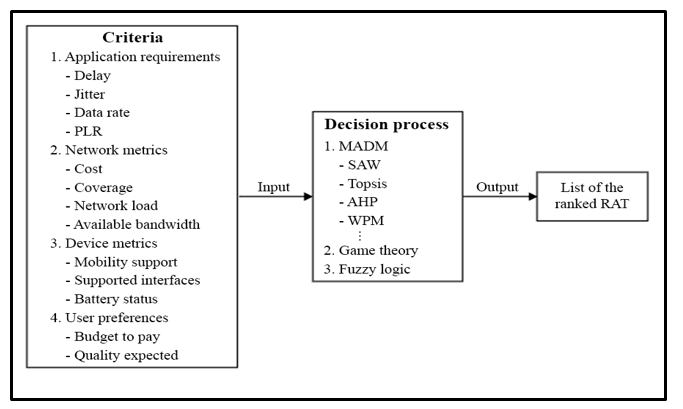}}
\caption{The access network selection procedure.\label{fig2} \cite{bendaoud2021multicriteria}}
\end{figure*}

 A fundamental component of heterogeneous wireless networks is the access network selection problem, which tries to give users optimum connectivity at all times. Handoffs are challenging for static models in real-world applications because of practical restrictions like user mobility that result in connection deterioration when switching between WIFI and mobile networks. In order to retain stability while reacting to current network conditions, an effective solution must be intelligent and dynamic.\\
 Thus, the purpose of the access network selection framework is to offer the user the proper RAT for an uninterrupted, smooth user experience. The well-known notion is QoS-based access network selection, where offering the user the greatest possible QoS is the top priority. Additionally, there are QoE-based models (Quality of Experience) where the aim is to attain the best QoE. However, given QoE is a subjective value, it is challenging to construct solid models for QoE-based network selection. Several criteria are used in this optimisation problem, see Fig \ref{fig2}\\
 Most techniques used in this context are MADM; Without presenting the mathematical models, we will merely provide an overview of several of these strategies.
\begin{itemize}
\item \textit {SAW Simple Additive Weight:} Its inputs are a matrix of data and a vector of weights; hence, for each row, it multiplies the data by its related weight and adds it all. The data has to be normalised first before conducting computations.
\item \textit {TOPSIS Technique for Order of Preference by Similarity to Ideal Solution:} TOPSIS rates the options according to their distance from the positive and negative ideal solutions; the appropriate choice is the closest to the ideal positive solution and the furthest from the negative ideal solution.
\item \textit{AHP Analytic Hierarchy Process:} It has a hierarchical structure, from aim to criterion to alternatives; it employs pairwise comparisons to establish the weight of qualities and ranks the alternatives.
\item \textit{WPM Weighted Product Method:}Its premise is the same as SAW technique; it employs a weighted product instead of addition.
\end{itemize}
\subsection{DQN and access network selection}
 Deep Q-network is a reinforcement learning approach that combines the Q-learning algorithm with deep neural networks; we will describe how to employ this idea in our scenario, where the access network selection is modelled as a Markov decision process. \\
 The concept of DQN is to approximate the Q-value function Q(s, a), which assesses the long-term benefit of performing action “a” on state “s”. 
 The DQN algorithm contains many critical components:
 \begin{itemize} 
 \item  The agent:  In our situation, it is the User Equipment UE that begins the rating process.
 \item  The environment: the collection of multiple choices, in our instance, the different networks (Wi-Fi, 4G (LTE), 5G (eMMB on mmWave), Satellite (LEO Starlink))
 \item  The states: QoS settings, network, and device metrics.
 A state vector is as follows:  \\
 $S_{(net_i)}=[Throughput_{(net_i )},latency_{(net_i)},jitter_{(net_i)},\\
 PLR_{(net_i )},U_{(net_i )},C_{(net_i)}] $, it reflects the status of a network at various instants.
 \item  The actions: picking the network to connect to.
 \item  The reward:  It is the feedback on connection quality; it may be utilised subsequently as a QoE-based method.
 \end{itemize}
The reward function reflects how the chosen action meets the optimisation goals.  It may be characterised in numerous ways.  The next text is devoted to our reward function.  Our purpose is to prefer a network with a larger bandwidth, with reduced latency, jitter, packet loss, not loaded, and with less expense. \\
 So, the basic notion is that the weights would be dynamic; they rely on the value of the parameter.  The typical form is $W = base + (X/scale)$, where the $base$ offers the initial significance, and it is also the minimum value of the weight.  $(X/scale)$, on the other hand, delivers a dynamic modulation based on the actual value of the parameter; it enables us to penalise the poor values and reward the great ones more.\\
So, the reward function is defined as\\
\begin{multline*}
R_{net_i} = 
\left[
W_{b(net_i)} \cdot \frac{b_{net_i}}{100}
- W_{l(net_i)} \cdot \frac{l_{net_i}}{300}
- W_{j(net_i)} \cdot \frac{j_{net_i}}{50}
\right. \\
\left.
- W_{p(net_i)} \cdot \frac{p_{net_i}}{10}
- W_{u(net_i)} \cdot \frac{u_{net_i}}{100}
- W_{c(net_i)} \cdot \frac{c_{net_i}}{10}
\right]
\end{multline*}
\begin{itemize} 
\item $W_b$ indicates the weight of the bandwidth, determined as \( W_b = 4 + \frac{b}{100} \).  So, the bigger the bandwidth, the higher \(W_b\).
\item $W_l$ reflects the weight of the delay, represented as \( W_l = 4 + \frac{l}{100} \).  greater latency leads to a greater \(W_l\) value.
\item $W_j$ is the weight of the jitter, given by \( W_j = 2.5 + \frac{j}{20} \).
\item $W_p$ is the weight of packet loss (together with the handover parameters), calculated as \( W_p = 4 + \frac{p}{5} \).  Higher packet loss is highly punished.
\item $W_u$ reflects network utilisation (congestion), defined as \( W_u = 3 + \frac{u}{50} \).  A heavily crowded network is less desirable.
\item $W_c$ indicates the cost for each network, defined as \( W_c = 2 + \frac{c}{10} \).
 \end{itemize}
This function enables us to normalise our reward, and we note that with a larger bandwidth, we get a bigger reward value, and it is inversely proportional to the delay, packet loss, and jitter, where a higher value penalises the network. \\
If we analyse our proposed reward function, we find that it has a positive partial derivative of bandwidth, which means that with an increase in the bandwidth, we get an increase in the reward function. For the other terms, they have a negative partial derivative, which implies that an increase in these terms decreases the reward function, and in both cases, it is coherent with reality. The function is monotone in all cases, whether it increases or decreases, because it is a linear function.\\
The architecture of the DQN is of three parts:
\begin{itemize} 
\item  Input layer: where the normalised data are input into the system.
\item  Hidden layers: a series of convolutional layers to capture complicated state-action interactions.
\item Output layer: Q-values with the reward for each option (network) that permits the ranking of networks.
\end{itemize}
The learning process involves multiple phases as follows:

  \begin{itemize} 
 \item  Exploitation and exploration: we are utilising the $\epsilon$-greedy method to balance between exploration, which implies looking for new networks to pick, and exploitation, which mandates exploiting the preceding findings (networks).
 \item Experience replay: storing previous data (actions, states, rewards, and upcoming states) and sampling it to decorrelate the data throughout the training, which increases the stability.
 \item Target network: a separate network used to calculate the Q-value.
 \item  The update of the Q value is dependent on the update of the Temporal Difference TD,
 \begin{center} 
 $Q(s_t,a_t )=Q(s_t,a_t )+\alpha[r_{(t+1)}+\gamma max{(a)}  Q(s_{(t+1)},a_t )-Q(s_t,a_t )] $ 
 \end{center} 
 \end{itemize}
$Q(s_t,a_t )$ is the Q-value of the state "s" and the action "a". \\ $\alpha $ The learning rate.\\ $r_{(t+1)} $ The reward for the action "a" on state s.\\ $\gamma $  The discount factor.\\ $max_{a} Q(s{(t+1)},a_t )$ indicates the maximum Q-value for all conceivable actions. \\
Since the proposal begins with an exploration, the algorithm will choose actions that may result in networks with poor rewards since it has little knowledge of the system and its surroundings. 
In order to avoid the local optimum, if we just choose the known actions, the goal is to test the various activities in an effort to discover the best new actions and strategies.  In order to balance exploration and exploitation, we are adopting a $\epsilon$-greedy strategy, with a probability of $\epsilon$ for exploration (i.e., selecting a random action) and a probability of $1-\epsilon$ for exploitation (i.e., selecting previously known actions). Additionally, $\epsilon$ is gradually degraded for each epoch (the $\epsilon$ decay procedure). The idea is that when $\epsilon$ is high, we are making more explorations, and as time goes on, we degrade the value of $\epsilon$ to balance and even make more exploitation when we consider that the earning has ended (we explored the maximum of different actions and we can know which action is good or not). \\
After conducting numerous tests and epochs, the algorithm's behaviour becomes evident and comprehensible. Initially, it produces average results (with poor decisions), but after a number of epochs, it becomes more accurate and produces good results. This indicates that the DQN has attained a high level of learning, and as we all know, making mistakes is a necessary part of learning. \\
The suggested DQN process for network selection is represented by algorithm \ref{algo11}. It's crucial to keep in mind that DQN is already an algorithm; we just modify it to fit our situation, which is access network selection.\\

\begin{algorithm}

\begin{algorithmic}
\State Initialize \texttt{step} $\gets 0$, \texttt{done} $\gets$ \textbf{False}
\While{not done}
    \State \texttt{step} $\gets$ \texttt{step} + 1
    \State Construct \texttt{statearray} from \texttt{fixedstate} as an array
    \If{rand(0,1) $<$ $\varepsilon$}
        \State \texttt{action} $\gets$ random choice from \{0, ..., \texttt{actionsize} - 1\} 
    \Else
        \State \texttt{qvalues} $\gets$ model.predict(\texttt{statearray})
        \State \texttt{action} $\gets$ argmax(\texttt{qvalues}) 
    \EndIf
    \State (\texttt{reward}, \texttt{done}, \texttt{networkname}) $\gets$ take\_action(\texttt{action}, \texttt{fixedstate})
    \State Construct \texttt{nextstatearray} from updated \texttt{fixedstate}
    \State Store transition: memory.append((\texttt{statearray}, \texttt{action}, \texttt{reward}, \texttt{nextstatearray}, \texttt{done}))
    \State Call \texttt{train\_dqn()} to update model
    \State \texttt{selectednetwork} $\gets$ \texttt{action}
\EndWhile
\State Increment \texttt{selectioncount[selectednetwork]}
\State Print episode result: selected network name and reward
\If{$\varepsilon > \varepsilon_{\text{min}}$}
    \State $\varepsilon \gets \varepsilon \times \varepsilon_{\text{decay}}$ 
\EndIf
\end{algorithmic}
\caption{DQN Network Selection}
\label{algo11} 
\end{algorithm}
The discount factor, the exploration rate, the learning rate, and the memory size are initialised at the start of Algorithm \ref{algo11}.  Additionally, it uses random weights to initialise the Q-network. \\ 
In order to train the model with fresh and varied QoS values, it then creates new QoS metrics at each epoch. Depending on the value of $\epsilon$, the first step in the learning process is to choose an action 'a' with exploration or exploitation.  After that, it performs action 'a', observes reward 'r', moves to state 's', and stores the whole process in 'D'. \\
We update the Q-network after calculating the goal 'y'.  A maintenance operation (updating the target network) will be applied for every "C" step.  Lastly, we decay $\epsilon$.

\section{Results and discussion}
\begin{figure*}[t!]
\centerline{\includegraphics[width=12cm, height=6cm]{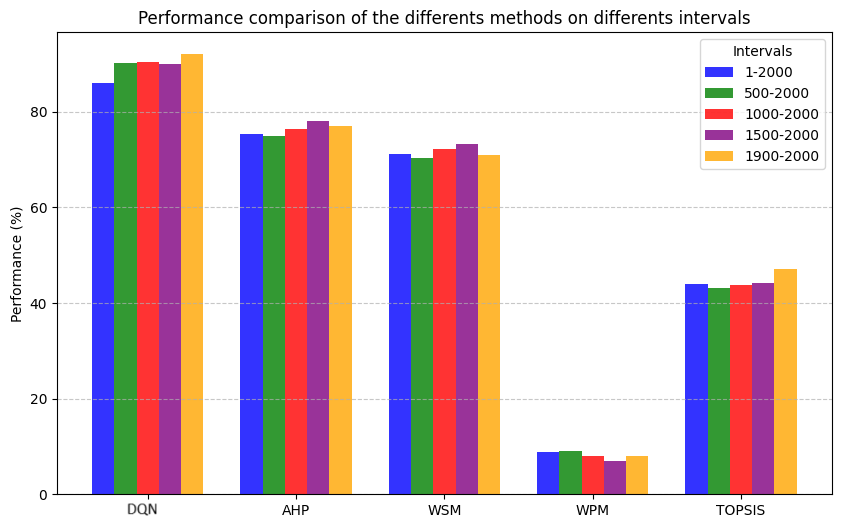}}
\caption{Performance comparison of the selection of the 5G of the different methods on different intervals of epochs.\label{fig4}}
\end{figure*}
\begin{figure*}[h]
\centerline{\includegraphics[width=12cm, height=6cm]{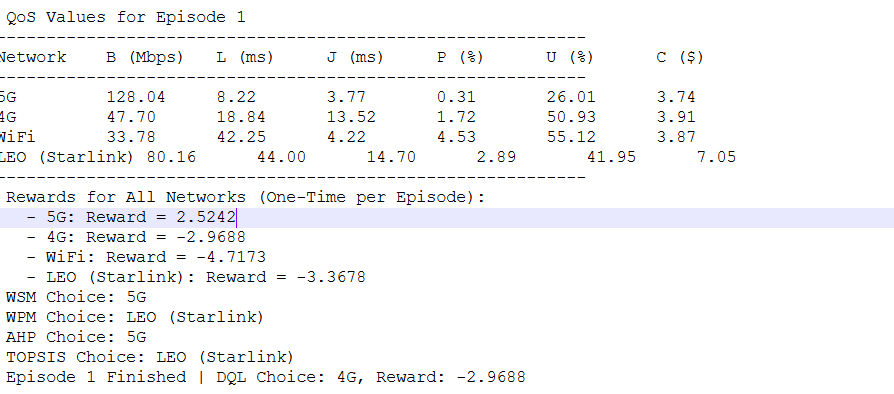}}
\caption{Trace file of the first epoch.\label{fig3}}
\end{figure*}

A table listing the various parameters shown in Table \ref{tab1} serves as a starting point for the simulation sequence. These values are produced by running simulations on NS3 in a variety of circumstances and scenario settings. The QoS values for the various networks are shown in Table \ref{tab1}.
\begin{table}[h!]
\centering
\caption{Network Characteristics}
\label{tab1}
\begin{tabular}{|p{3.5cm}|p{2cm}|p{2cm}|p{2cm}|p{2cm}|}
\toprule
\textbf{Metric} & \textbf{5G} & \textbf{4G} & \textbf{WiFi} & \textbf{LEO (Starlink)} \\ \midrule
Bandwidth (B) Mbps      & 50 -- 500  & 10 -- 50   & 20 -- 80   & 50 -- 200    \\
Latency (L) ms          & 5 -- 10    & 10 -- 30   & 10 -- 50   & 30 -- 70     \\
Jitter (J) ms           & 1 -- 5     & 5 -- 15    & 1 -- 8     & 5 -- 20      \\
Packet Loss (P) \%      & 0 -- 1     & 0.1 -- 2   & 0 -- 5     & 2 -- 10      \\
User Load (U) \%        & 10 -- 50   & 30 -- 70   & 20 -- 60   & 40 -- 80     \\
Cost (C) \$             & 3 -- 6     & 2 -- 5     & 1 -- 4     & 4 -- 8       \\ \bottomrule
\end{tabular}
\end{table}
In the DQN method, the actual entries are tables that have been created based on the Table \ref{tab1}, and subsequently, we produce fresh data for each epoch that can be used for learning purposes. For instance, the first epoch that can be seen in our trace file is shown in Figure \ref{fig3}. We have an entry for the various networks, as can be seen, and we have put into practice the different approaches that were taken into consideration in this work for the purpose of selecting an access network.  This example demonstrates that the conventional legacy techniques have chosen LEO Starlink, but DQN has opted for Wi-Fi with a negative return. As a result, our statements from the preceding paragraphs on exploration and exploitation are validated.
The first epochs are devoted to exploration; the DQN process is attempting alternative strategies and actions to explore the system; then, after a given number of epochs, it will learn and be able to choose the better actions and avoid the bad ones. We can immediately detect that 5G is superior to 4G in the situation described in Figure \ref{fig3}, WSM and AHP were right and selected the best RAT (5G), and our proposal DQN does not, it selected 4G, this is a typical behaviour since we are in the exploration phase. \\
So, as we said, we performed simulations until 2000 epochs (hardware limit), The results are presented below.\\
\begin{table}[h]
\centering
\caption{Percentage of 5G selection of the different methods on different intervals of epochs}
\label{tab11}
\begin{tabular}{|p{2cm}|p{2cm}|p{2cm}|p{2cm}|p{2cm}|p{2cm}|}
\toprule
Nb of epochs & {DQN} & {AHP} & {WSM} & {WPM} & {TOPSIS} \\
\midrule
\textbf{1--500}    & 74.08 & \textbf{77.2} & 74.2 & 12.4 & 50.4 \\
\textbf{500--1000} & \textbf{91.3}  & 72.8 & 67.0 & 17.1 & 41.8 \\
\textbf{1000--1500} & \textbf{92.5}  & 74.8 & 71.2 & 15.0 & 43.4 \\
\textbf{1500--2000} & \textbf{90.9}  & 78.2 & 73.4 & 13.7 & 44.2 \\
\textbf{1900--2000} & \textbf{93.0}  & 78.0 & 72.0 & 17.0 & 47.0 \\
\midrule
\textbf{1--2000}    & \textbf{87.03} & 75.4 & 71.2 & 15.3 & 44.05\\
\bottomrule
\end{tabular}
\end{table}

Figure \ref{fig4} pertains to the selection rate of the 5G network, since it is the most probable optimal network for all entries. The results indicate that, with the exception of the initial 500 epochs (characterised by greater exploration than exploitation), DQN is generally superior across all epoch intervals. Furthermore, a detailed analysis reveals that as exploration diminishes, DQN's performance improves correspondingly, which is logical, as the system has identified the optimal actions and ceases to engage in exploration of unknown actions.
The average selection of 5G for the DQN is about 87\%, but the maximum number for the AHP, the other MADM approach, is 75.5\%, indicating a significant disparity. Observe that DQN diminishes during the first 500 epochs of investigation; thereafter, the percentages exceed 90\%. This unequivocally demonstrates that the DQN learns well and selects optimal actions.\\
It is vital to mention that we were meant to mimic our idea on multiple type of applications: VoIP, video streaming, video calls, and other services like web browsing and file downloads. But we are now 100\% positive that it will deliver identical outcomes since the power of the proposal is built mainly on the learning process, which is the same in all settings.  We are certain that the first phase of exploitation will provide moderate performance from the DQN; thereafter, the proposal will surpass all other solutions as the process advances in exploitation, which is the proposal's key strength, "the learning feature".
\section{Conclusion}
A novel and innovative method for selecting an access network in a heterogeneous wireless environment is described. Reinforcement learning, and specifically the Deep Q-Network algorithm (DQN), was used. Although DQN's resource-intensive methodology made it difficult to utilise it in the network selection problem, once trained, it produces much better results than MADM, a standard static technique that attempts to rank items each time without taking into account the rankings of the preceding iteration. Compared to the MADM approaches, we first saw that the algorithm was performing poorly throughout the learning phase. However, after a specific number of epochs, the algorithm began to significantly outperform the other methods. The real-time aspect of this work is still its limit; we must train our algorithm with a large amount of data, which implies that the more epochs, the better the performance. Therefore, from this point of view, it is crucial to implement this solution on numerous powerful machines that enable us to go beyond 2000 epochs, and we are confident that the accuracy will increase more and more.

%
%
 \bibliographystyle{splncs04}
 \bibliography{bibtex}
%
%
%
%
%
\end{document}